\documentclass{PoS}
\usepackage{amssymb,fontenc,times,mathptmx}

\usepackage{psfrag}
\usepackage{slashed}
\usepackage{cancel}
\usepackage{lscape}
\usepackage{caption}
\usepackage{array}
\usepackage{graphicx}

\usepackage{cite}

\usepackage[utf8]{inputenc}

\definecolor{mygreen}{rgb}{0,0.7,0}

\definecolor{myblue}{rgb}{0,0,0.7}

\definecolor{myred}{rgb}{0.7,0,0}

\def\braket#1{\langle #1 \rangle}

\def\usegraph#1#2{\includegraphics[scale=1.0,trim=0 #1 0 0]{graphs/#2.pdf}}
\def\usexfig#1#2#3{\includegraphics[scale=#3,trim=0 #1 0 0]{xfigs/#2.pdf}}

\usepackage{graphicx}
\usepackage{amsmath}

\def\cN{\mathcal{N}}

\def\cQ{\mathcal{Q}}

\def\k{\ell}
\def\d{\mathrm{d}}

\def\la{\langle}
\def\ra{\rangle}
\def\spA#1#2{\la#1#2\ra}

\DeclareMathOperator{\tr}{\rm tr}

\def\trp{\tr_+}

\def\eps{\epsilon}

\title{Local integrands for two-loop QCD amplitudes}

\ShortTitle{Local integrands in QCD}

\author{\speaker{Simon Badger}\\
        Higgs Centre for Theoretical Physics, School of Physics and Astronomy, The University of Edinburgh, Edinburgh EH9 3JZ, Scotland, UK\\
        E-mail: \email{simon.badger@ed.ac.uk}}

\author{Gustav Mogull\\
        Higgs Centre for Theoretical Physics, School of Physics and Astronomy, The University of Edinburgh, Edinburgh EH9 3JZ, Scotland, UK\\
        E-mail: \email{g.mogull@ed.ac.uk}}

\author{Tiziano Peraro\\
        Higgs Centre for Theoretical Physics, School of Physics and Astronomy, The University of Edinburgh, Edinburgh EH9 3JZ, Scotland, UK\\
        E-mail: \email{tiziano.peraro@ed.ac.uk}}

\abstract{In this talk we review the recent computation of the five- and six-gluon two-loop
amplitudes in Yang-Mills theory using local integrands which make the infrared
pole structure manifest. We make some remarks on the connection with BCJ relations
and the all-multiplicity structure.}

\FullConference{Loops and Legs in Quantum Field Theory\\
		24-29 April 2016\\
		Leipzig, Germany}

\begin{document}

\section{Introduction}

Since the last edition of Loops and Legs in Quantum Field Theory, there has been rapid progress in
computing NNLO QCD corrections to differential distributions of $2\to 2$ scattering processes relevant for
on-going experiments at the LHC. While there has been great progress in analytic loop integration methods, this
recent success has been largely thanks to new methods for subtracting infrared singularities. At this
conference we have seen a number of recent highlights and new results presented~\cite{Boughezal:2016wmq,Ridder:2016nkl,Currie:2016ytq,%
DelDuca:2016ily,Czakon:2016dgf,Hoff:2016bdo,Borowka:2016ehy,Grazzini:2016swo,Grazzini:2016ctr}. For a more complete review
of the current state of the art we refer the reader to ref. \cite{Badger:2016bpw}.

For higher-multiplicity processes beyond $2\to 2$ scattering, which the LHC experiments will be increasingly
sensitive to, little is known: the bottleneck continues to be computation of two-loop amplitudes.
Attempts to alleviate this bottleneck, in the spirit of the solution to the one-loop case, through
automation with algebraic algorithms have achieved some success~\cite{Mastrolia:2011pr,Badger:2012dp,Zhang:2012ce,Mastrolia:2012an,%
Badger:2012dv,Kleiss:2012yv,Feng:2012bm,Mastrolia:2012wf,Mastrolia:2013kca,Mastrolia:2016dhn,Kosower:2011ty,Larsen:2012sx,%
CaronHuot:2012ab,Johansson:2012zv,Johansson:2013sda,Johansson:2015ava,Sogaard:2013fpa,Sogaard:2013yga,Sogaard:2014jla,Sogaard:2014oka,%
Sogaard:2014ila}. Though there are still obstacles
to overcome before this approach can be applied to realistic phenomenological studies, early results
for $2\to3$ scattering amplitudes have been obtained using the on-shell technique of $D$-dimensional
generalised unitarity together with integrand reduction~\cite{Badger:2013gxa,Badger:2015lda}. Another important step in this direction has been
the completion of the planar master integrals
for $2\to3$ scattering both in the fully massless~\cite{Gehrmann:2015bfy,Papadopoulos:2015jft} and single off-shell leg~\cite{Papadopoulos:2015jft}
cases.

In this contribution we investigate how local integrands can be used to simplify analytic representations of multi-leg two-loop
amplitudes in pure Yang-Mills theory. The main principle is to build an integrand representation in which infrared singularities are manifest and a basis of
simple integrals can be identified. Originally developed for use in $\cN=4$ maximally supersymmetric Yang-Mills (sYM) theory, these local integrands have been successfully used to obtain results in many high-multiplicity two- and three-loop cases~\cite{ArkaniHamed:2010kv,ArkaniHamed:2010gh,Bourjaily:2015jna}.
Recently, we have found such a local representation for the five- and six-gluon amplitudes in Yang-Mills with all positive helicities~\cite{Badger:2016ozq}.
The all-plus sector of Yang-Mills theory is a useful testing ground for new methods since the two-loop amplitudes are the simplest in a non-supersymmetric theory and they are closely related to those the $\cN=4$ sYM.

Since the amplitudes vanish at tree level they are also amenable to one-loop techniques. A combination
of four-dimensional unitarity cuts and augmented BCFW recursions has recently been used to obtain expressions for the five- and six-gluon
amplitudes of interest here~\cite{Dunbar:2016aux,Dunbar:2016cxp,Dunbar:2016gjb}. In this work we test the two-loop
technology of $D$-dimensional unitarity using six-dimensional tree input to obtain $D$-dimensional integrand representations
of the these amplitudes which are free of all spurious singularities.

\section{Generalised unitarity and integrand reduction}

Building on the techniques developed for computing one-loop amplitudes,
generalised unitarity cuts and integrand-level parametrisations can be
combined into a purely algebraic procedure for evaluating multi-loop integrands
from tree-level amplitudes. At the integrand level, a colour-ordered two-loop
amplitude can be decomposed into a set of topologies with irreducible
numerators,
\begin{align}
  \hat{A}^{(2)} = \sum_{{\rm topologies}\,\,T} I^{4-2\epsilon}\left[\Delta(T)\right],
\end{align}
where $\hat{A}$ is the amplitude normalised to the Parke-Taylor factor,
\begin{equation}
  \hat{A} = A \prod_{i=1}^n \braket{i,i+1}
\end{equation}
and
\begin{align}
I^{4-2\epsilon}[\Delta(T)]
\equiv-(4\pi)^{4-2\epsilon}\int\frac{\d^{4-2\epsilon}\ell_1\d^{4-2\epsilon}\ell_2}{(2\pi)^{2(4-2\epsilon)}}
\frac{\Delta(T)}{\prod_{\alpha\in T}\cQ_\alpha(p_i;\ell_1,\ell_2)}.
\end{align}
for a set of loop propagators $1/\cQ_\alpha$. In the following we will use the graph of
each topology to represent the particular set of propagators appearing.

An irreducible numerator $\Delta$ is a polynomial in
a set irreducible scalar products (ISPs),
defined by polynomial division with respect to a Gr\"obner basis of the
propagators~\cite{Zhang:2012ce,Mastrolia:2012an},
\begin{align}
  P^{(\vec{r})}({\rm ISP}) \,/\, {\rm Gr}(\{\cQ_\alpha({\rm ISP})\}) = \Delta({\rm ISP}),
\end{align}
where $\vec{r}$ is a set of restrictions on the maximum rank of each loop momentum
set by the renormalizability of the theory under consideration.

Rational coefficients of these ISPs can be extracted from
$D$-dimensional generalised unitarity cuts.
On each multiple cut the amplitude factorises
into a product of tree-level amplitudes: at two loops,
cuts must be performed in a minimum of six dimensions to obtain complete, dimensionally regulated amplitudes.
We therefore describe kinematic variables using the six-dimensional
spinor-helicity formalism of Cheung and O'Connell~\cite{Cheung:2009dc}.
To control kinematic complexity,
these six-dimensional spinors are constructed from the external four-dimensional
momenta using a rational parametrisation in terms of
Hodges' momentum twistors~\cite{Hodges:2009hk}.
Further details of the explicit parametrisation of the kinematics
used can be found in ref.~\cite{Badger:2016uuq}.

After performing this integrand reduction there remains
an additional redundancy in the basis of integrals.
These additional simplifications can be identified by solving
systems of integration-by-parts (IBP) identities.
There is an on-going effort to improve the efficiency of these algorithms for use
in the context of an algebraic reduction procedure which hopes to alleviate
this issue~\cite{Gluza:2010ws,Schabinger:2011dz,Ita:2015tya,Larsen:2015ped}.
Nevertheless,
there remains a remarkable flexibility in the basis of ISP polynomials.
ISP choices that make physical properties of the amplitude manifest
can lead to dramatic simplifications.

A particularly powerful technique in $\cN=4$ sYM has been the use of local
integrands to control infrared singularities and provide a basis of simple integrals~\cite{ArkaniHamed:2010kv,ArkaniHamed:2010gh,Bourjaily:2015jna}.
In this work we have explored the use of these numerator functions in
dimensionally-regulated amplitudes outside four-dimensional $\cN=4$.

\section{Local integrands for one-loop amplitudes}

The simplest local numerator can be seen in the
the one-loop box topology with all massless external legs.
The commonest integral basis uses the scalar box function
\begin{align}\label{eq:boxint}
I^{4-2\eps}\bigg(\usegraph{10}{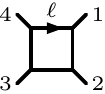}\bigg)
=-i(4\pi)^{2-\eps}\int\!\frac{\d^{4-2\eps}\k}{(2\pi)^{4-2\eps}}
\frac{1}{\k^2(\k-p_1)^2(\k-p_{12})^2(\k+p_4)^2}.
\end{align}
where $p_{i\cdots j}=p_i+p_{i+1}+\cdots+p_j$.
This integral has singularities when the loop momentum flowing through any propagator becomes soft
or when it becomes collinear to either two of the external momenta adjacent to it.

A numerator function
\begin{align}
\trp(1(\k\!-\!p_1)(\k\!-\!p_{12})3)
=\tfrac{1}{2}\tr((1+\gamma_5)p_1(\k\!-\!p_1)(\k\!-\!p_{12})p_3)
\end{align}
can be used to regulate the integral as it vanishes in these singular regions. Following the notation introduced by Arkani-Hamed and collaborators,
we denote this numerator as a wavy line across the box joining legs 1 and 3,
\begin{align}\label{eq:localboxint}
I^{4-2\eps}\bigg(\usegraph{10}{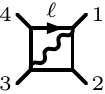}\bigg)
=-i(4\pi)^{2-\eps}\int\!\frac{\d^{4-2\eps}\k}{(2\pi)^{4-2\eps}}
\frac{\trp(1(\k\!-\!p_1)(\k\!-\!p_{12})3)}{\k^2(\k-p_1)^2(\k-p_{12})^2(\k+p_4)^2}.
\end{align}
This integral in four dimensions can now be identified as a
box integral in six dimensions:
\begin{align}
  I^{4-2\eps}\left( \usegraph{10}{boxs} \right)
  = (-1+2\eps) s_{13} I^{6-2\eps} \left( \usegraph{10}{box} \right)
  = -\frac{r_\Gamma}{2} \left(\log^2\left(\frac{s_{12}}{s_{23}}\right) + \pi^2\right)+ \mathcal{O(\eps)},
\end{align}
which is manifestly finite.

\section{Two-loop results for the all-plus amplitudes}

In a recent paper we studied the planar all-plus sector at two loops
from the perspective of local numerators and found compact
representations for the five- and six-gluon cases~\cite{Badger:2016ozq}.
Rather than reproducing any of the local-integrand based planar results here,
we will make some remarks on a possible all-multiplicity structure
that agrees with these existing results.
Though there are still a few unknowns quantities,
there are certainly some patterns emerging that lead us to make conjectures
about the general $n$-point case.
This work follows the original derivation of the five-point numerators
\cite{Badger:2013gxa} and the recent calculation of the nonplanar
contributions \cite{Badger:2015lda}.

The simplicity of the all-plus sector of Yang-Mills theory is made apparent
by its somewhat mysterious relation to $\cN=4$ sYM.
At one-loop order this amounts to a dimension-shifting relation~\cite{Bern:1996ja},
expressible at the integrand level as
\begin{align}
  \hat{A}^{(1)}(1^+,2^+,\cdots,n^+)
=\frac{i(D_s-2)}{(4\pi)^{2-\eps}\spA{1}{2}^4}
\sum_{T\in\text{boxes}}
I^{4-2\epsilon}\left[\mu^4\Delta^{[\cN=4]}_{--+\dots+}(T)\right],
\end{align}
where the one-loop integration operator is the same as
that used in eq.~(\ref{eq:boxint}), $\mu^2=-(\k^{[-2\eps]})^2$
and $D_s$ is the spin dimension of internal gluons.

At two-loop order we draw a similar correspondence with the
local-integrand presentation of the $\cN=4$ amplitude given in
refs.~\cite{ArkaniHamed:2010kv,ArkaniHamed:2010gh}.
This takes the form
\begin{align}
\frac{A^{(2),[\cN=4]}(1,2,\cdots,n)}{A^{(0),[\cN=4]}(1,2,\cdots,n)}
=\frac{i}{2(4\pi)^{4-2\epsilon}}
\sum_{a<b<c<d<a}\frac{\braket{abcd}}{\braket{ab}\braket{cd}}
I^{4-2\epsilon}\bigg(\usexfig{110}{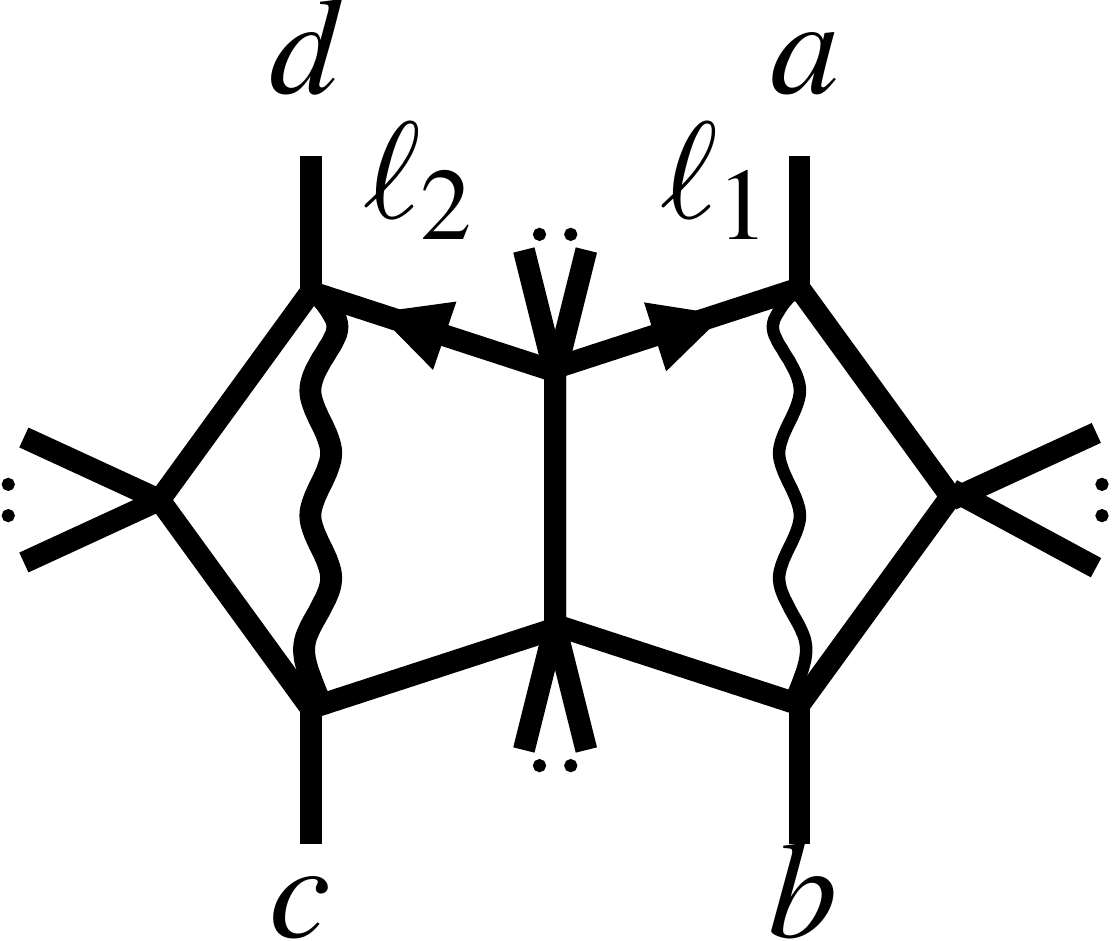}{.15}\bigg),
\end{align}
where the notation for wavy lines is the same as
that used in eq.~(\ref{eq:localboxint}).
The sum is over cyclically-ordered labels $a$, $b$, $c$, $d$,
the cases $b=a+1$, $c=b+1$, $d=c+1$ and $a=d+1$ corresponding to
various additional clustered legs around the diagram disappearing.
We have also introduced the totally antisymmetric twistor four bracket,
which can be written explicitly in terms of spinor variables as
\begin{align}
\braket{ijkl}=\braket{ij}\braket{kl}
\left(p_{i+1,k}^\mu-\frac{\braket{i|\gamma^\mu p_{i+1,j}|j}}{2\braket{ij}}
+\frac{\braket{k|\gamma^\mu p_{k+1,l}|l}}{2\braket{kl}}\right)^2.
\end{align}

In the all-plus sector at two loops we make use of the
extra-dimensional ISP polynomials
\begin{align}
  F_1(\ell_1^{[-2\epsilon]},\ell_2^{[-2\epsilon]}) &= (D_s-2)(\mu_{11}\mu_{22}\!+\!(\mu_{11}\!+\!\mu_{22})^2\!
+\!2\mu_{12}(\mu_{11}\!+\!\mu_{22}))\!+\!16(\mu_{12}^2\!-\!\mu_{11}\mu_{22}), \\
  F_2(\ell_1^{[-2\epsilon]},\ell_2^{[-2\epsilon]}) &= (D_s-2) (\mu_{11}+\mu_{22})\mu_{12}, \\
  F_3(\ell_1^{[-2\epsilon]},\ell_2^{[-2\epsilon]}) &= (D_s-2)^2 \mu_{11}\mu_{22},
\end{align}
where we have introduced the ISPs $\mu_{ij}=-\k_i^{[-2\eps]}\cdot\k_j^{[-2\eps]}$.
The full amplitude is
\begin{align}
  \hat{A}^{(2)}(1^+,2^+,\cdots,n^+)
=-\frac{i}{2(4\pi)^{4-2\epsilon}}
\sum\,\,I^{4-2\epsilon}\left[\Delta\bigg(\usexfig{135}{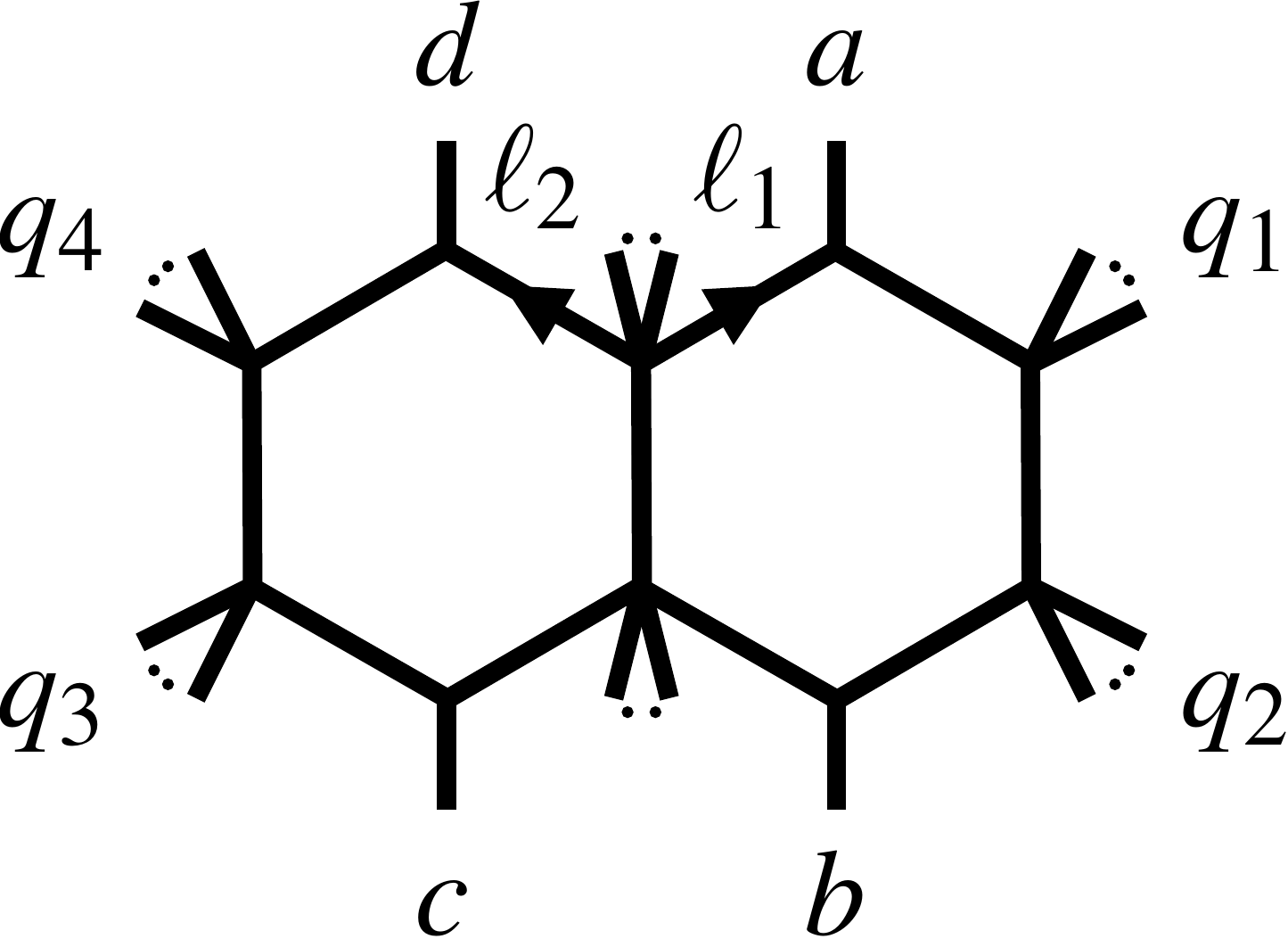}{.15}\bigg)
+\rule{0cm}{1cm}\Delta\bigg(\usexfig{115}{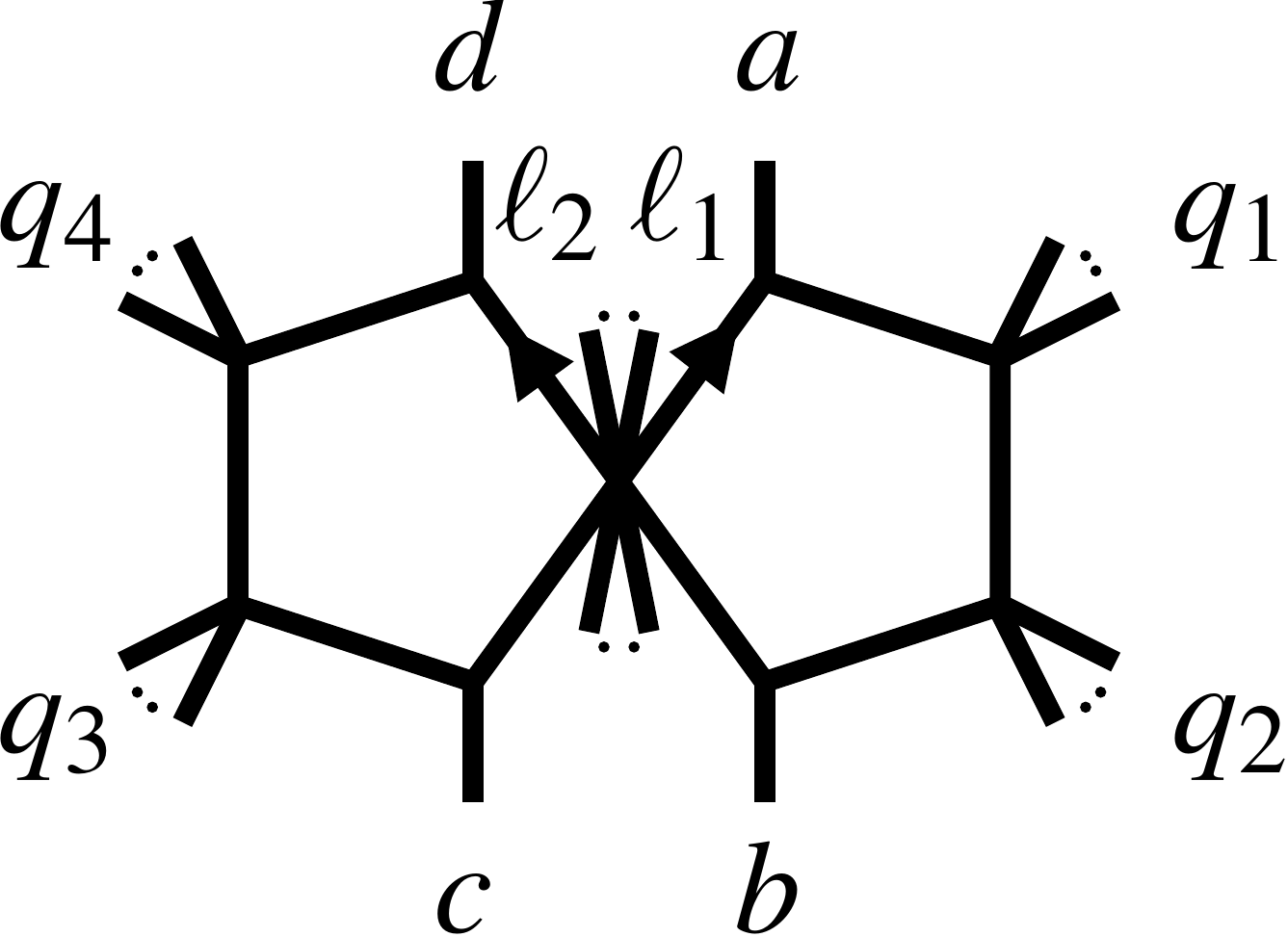}{.15}\bigg)\right],
\end{align}
where now, in addition to ways of allocating the legs $a$, $b$, $c$, $d$,
we also sum on inequivalent ways of distributing the other
massless legs between the $q_i$.
All genuine two-loop topologies have numerators
\begin{align}\label{eq:twoloopdelta}
\Delta\bigg(\usexfig{135}{551}{.15}\bigg) &=
  -\frac{\braket{abcd}}{\braket{ab}\braket{cd}}
  N_{ab}(q_1,q_2;\ell_1)N_{dc}(q_4,q_3;\ell_2) F_1, 
\end{align}
while the one-loop-squared numerators are
\begin{align}
\Delta\bigg(\usexfig{115}{550}{.15}\bigg)
&=\frac{\trp(abcd)}{s_{ab}s_{cd}}N_{ab}(q_1,q_2;\ell_1)N_{dc}(q_4,q_3;\ell_2)
\left(F_2+F_3\chi_{abcd}(p_i;\ell_1,\ell_2)\right).
\end{align}
The functions $N_{ab}$ in various cases are
\begin{align}
N_{ab}(p,q;\ell)&=-\trp(p_apqp_b)\mu^2,\\
N_{ab}(p,0;\ell)&=\trp(p_a(\ell\!-\!p_a)(\ell\!-\!p_a\!-\!p)p_b)(\ell-p_a-p)^2,\\
N_{ab}(0,q;\ell)&=\trp(p_a(\ell\!-\!p_a)(\ell\!-\!p_a\!-\!q)p_b)(\ell-p_a)^2,\\
N_{ab}(0,0;\ell)&=s_{ab}(\ell-p_a)^4.
\end{align}
We see that all-plus graphs with supersymmetric counterparts
share the same numerators,
the all-plus versions carrying the additional function $F_1$.

Though we do not have a general expression for $\chi_{abcd}$,
we do know that
\begin{align}\label{eq:chidef}
\chi_{a,b,b+1,a-1}(p_i;\ell_1,\ell_2)
=\frac{p_{ab}^2+(\ell_1+\ell_2)^2}{p_{ab}^2},
\end{align}
corresponding to the particular case when a four-point vertex
lies at the center of the graph.
A partial explanation for this phenomenon can be found from BCJ tree-amplitude
identities~\cite{Bern:2008qj} as applied to the generalised unitarity cuts,
which will be explored in the next section.

Another beneficial property of this representation is that the universal infrared
pole structure can be identified directly at the integrand level as a virtue
of the numerators that control the soft and collinear singluarities. Further examination
of the soft regions of the integrals appearing has been presented in ref.~\cite{Badger:2016ozq}.

\section{Connection to BCJ tree-amplitude identities}

As mentioned previously, the common structure of one-loop-squared graphs whose maximal cuts contain a central four-gluon amplitude can, to some extent,
be explained by considering BCJ tree-amplitude identities.
Taking the box-triangle as an example,
the relevant maximal cut can be decomposed into irreducible numerators as
\begin{align}
\text{Cut}\bigg(\usegraph{13}{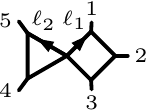}\!\bigg)
=\frac{1}{\braket{12}\braket{23}\braket{34}\braket{45}\braket{51}}\left(
\Delta\bigg(\usegraph{13}{430}\!\bigg)+\frac{1}{(\ell_1+\ell_2)^2}\Delta\bigg(\usegraph{9}{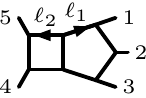}\bigg)\right),
\end{align}
where all exposed propagators are cut.
When understood as a sum over internal helicities of products of tree-level amplitudes,
this cut also satisfies a BCJ tree-amplitude relation
\begin{align}\label{eq:bcjcut}
(\ell_1+\ell_2)^2\text{Cut}\bigg(\usegraph{13}{430}\!\bigg)
=(\ell_1+p_{45}-\ell_2)^2\text{Cut}\bigg(\usegraph{13}{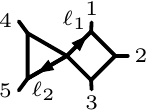}\!\bigg),
\end{align}
which is merely another way of writing $s_{12}A^{(0)}(1,2,3,4)=s_{13}A^{(0)}(1,3,2,4)$.
The most general polynomial solution for the box triangle on its cut is therefore
\begin{align}
s_{45}\Delta\bigg(\usegraph{13}{430}\!\bigg)
=\Delta\bigg(\usegraph{9}{431}\bigg)
-\frac{\braket{34}\braket{51}}{\braket{35}\braket{41}}\Delta\bigg(\usegraph{9}{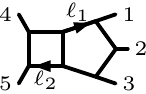}\bigg)
+(s_{45}+(\ell_1+\ell_2)^2)X,
\end{align}
which is a kinematic Jacobi identity.
Extra angle brackets appear as the numerators $\Delta(T)$
exclude the Parke-Taylor factor,
and we have introduced an unknown polynomial $X$ of external and loop momenta
as the numerators do not satisfy colour-kinematics duality.

Using the general formula (\ref{eq:twoloopdelta}) for the pentabox numerator,
we find that
\begin{align}
\Delta\bigg(\usegraph{13}{430}\!\bigg)
=\trp(1(\ell_1\!-\!p_1)(\ell_1\!-\!p_{12})345)F_2
+\frac{s_{45}+(\ell_1+\ell_2)^2}{s_{45}}X,
\end{align}
where we have used $F_2(\ell_1^{[-2\epsilon]},\ell_2^{[-2\epsilon]})
=F_1(\ell_1^{[-2\epsilon]},\ell_2^{[-2\epsilon]})-F_1(\ell_1^{[-2\epsilon]},-\ell_2^{[-2\epsilon]})$.
We can use this form to motivate eq.~(\ref{eq:chidef}) and use the explicit cut to
check that $X = F_3\trp(1(\ell_1\!-\!p_1)(\ell_1\!-\!p_{12})345)$.
Since we only required the BCJ relation for the central four-point amplitude this argument
is straightforward to apply to any topology of the type in eq.~(\ref{eq:chidef}).

For one-loop-squared cuts containing a central five- or six-gluon amplitude
the BCJ tree-amplitude relation is less simple, which is reflected in the numerators being less compact.
The BCJ procedure used here is the same as that used to compute nonplanar numerators 
from their planar counterparts in ref.~\cite{Badger:2015lda};
here we restrict ourselves to the planar sector and gain incomplete information.

\section{Outlook}

In these proceedings we have summarised our recent attempts at finding
compact analytic expressions for multi-leg two-loop amplitudes
free of spurious singularities.
An extension to arbitrary helicity configurations is clearly the most
important step before any phenomenological applications can be considered.
Though there are many unanswered questions in how a complete basis
of local integrands might be found for these cases,
we hope the ideas presented here will be a useful guide on how to proceed.
It would interesting to see how the recent understanding of non-planar corrections
in $\cN=4$~\cite{Bern:2015ple} applies to the Yang-Mills case which are a needed for a full
colour description of the amplitudes.

\section*{Acknowledgments}
We would like to thank Donal O'Connell for useful conversations and the organisers of Loops and Legs for another stimulating conference.
S.B. is supported by an STFC Rutherford Fellowship ST/L004925/1 and T.P. is supported by Rutherford Grant ST/M004104/1.  G.M. is supported by an STFC Studentship ST/K501980/1.


\providecommand{\href}[2]{#2}\begingroup\raggedright\endgroup

\end{document}